\documentclass[reprint,
frontmatterverbose, 
showpacs,preprintnumbers,
nofootinbib,
showkeys, 
aps,
prd,
longbibliography,
floatfix
]{revtex4-1}

\usepackage{amssymb}
\usepackage{amsmath}
\usepackage[american]{babel}
\usepackage{booktabs}
\usepackage[T1]{fontenc}
\usepackage[utf8]{inputenc}
\usepackage{txfonts}
\usepackage{url}
\usepackage[dvipsnames]{xcolor}
\usepackage{float}
\usepackage{graphicx}
\usepackage{dcolumn}
\usepackage{bm}
\usepackage{hyperref}
\usepackage[mathlines]{lineno}

\usepackage{siunitx}
\usepackage{hhline}
\usepackage{multirow}
\usepackage[dvipsnames]{xcolor}

\hypersetup{linktocpage=true,colorlinks=true,breaklinks,
linkcolor=[rgb]{0.0,0.0,1.0},filecolor=cyan,citecolor=[rgb]{0.8,0.00,0.0},urlcolor=[rgb]{0.79,0.08,0.48}}

\usepackage{color, colortbl}
\definecolor{LightCyan}{rgb}{0.88,1,1}
\definecolor{Gray}{gray}{0.9}

\begin{document}

\title{A More Accurate and Competitive Estimative of $H_0$ in Intermediate Redshifts.}
\author{G. Pordeus-da-Silva}
\email{givalpordeus@ufrn.edu.br}
\author{A. G. Cavalcanti}
\email{a.gcavalcanti@hotmail.com}
\affiliation{Departamento de F\'isica Te\'orica e Experimental, Universidade
	Federal do Rio Grande do Norte.\\ Campus Universit\'{a}rio,
	s/n - Lagoa Nova - CEP 59072-970, Natal, Rio Grande do Norte, Brasil.
}

%\author{G. Pordeus da Silva}
%\email{givalpordeus@ufrn.edu.br}

%\affiliation{
%	Departamento de F\'isica Te\'orica e Experimental, Universidade
%	Federal do Rio Grande do Norte.\\ Campus Universit\'{a}rio,
%	s/n - Lagoa Nova - CEP 59072-970, Natal, Rio Grande do Norte, Brasil.
%}
\date{\today }

\begin{abstract}
In order to clarify the tension between estimates of the Hubble Constant ($H_0$) from local ($z \ll 1$) and global ($z \gg 1$) measurements, Lima and Cunha (LC) proposed a new method to measure $H_0$ in intermediate redshifts ($z \approx 1$), which were obtained $H_0 = 74.1 \pm 2.2$ km s$^{-1} $Mpc$^{-1}$ ($1 \sigma$), in full agreement to local measurements via Supernovae/Cepheid dataset. However, Holanda \textit{et al.} (2014) affirm that a better understanding of the morphology of galaxy clusters in LC framework is needed to a more robust and accurate determination of $H_0$. Moreover, that kind of sample has been strongly questioned in the literature. In this context, (i) we investigated if the sample of galaxy clusters used by LC has a relevant role in their results, then (ii) we perform a more accurate and competitive determination of $H_0$ in intermediate redshifts, free of unknown systematic uncertainties. First, we found that the exclusion of the sample of galaxy clusters from the determination initially proposed by LC leads to significantly different results. Finally, we performed a new determination in $H_0$, where we obtained $H_0 = 68.00 \pm 2.20$ km s$^{-1}$ Mpc$^{-1}$ ($1 \sigma$) with statistical and systematic errors and $H_0 = 68.71^{+1.37}_{-1.45}$ km s$^{-1}$ Mpc$^{-1}$ ($1 \sigma$) with statistical errors only. Contrary to those obtained by LC, these values are in full harmony with the global measurements via Cosmic Microwave Background (CMB) radiation and to the other recent estimates of $H_0$ in intermediate redshifts.

\begin{description}
\item[Journal reference]
\cite{daSilva2018}  G. Pordeus da Silva and A. G. Cavalcanti, \href{https://doi.org/10.1007/s13538-018-0581-9}{Brazilian Journal of Physics \textbf{48}, 521 (2018)}.
%\item[DOI]
%\href{https://doi.org/10.1007/s13538-018-0581-9}{https://doi.org/10.1007/s13538-018-0581-9}
\end{description}
\end{abstract}
%\pacs{Valid PACS appear here}
\maketitle

%\preprint{APS/123-QED}

%\affiliation{%


 %Use showkeys class option if keyword

%\tableofcontents

\section{Introdu\c{c}\~{a}o}
\label{intro}
Even with the remarkable advent of modern cosmology, the existing tension between the Hubble Constant measurements ($H_0$) via local (via SNe Ia and Cepheids data) and global (via CMB data) measurements has been intensified in recent years. A significant and persistent tension between those two measurements may suggest evidence for a new fundamental physics beyond the standard model and for General Relativity or non-zero curvature \cite{Riess2016}. Therefore, the comparison between measurements of $H_0$ raised from different methods provides a test for the standard cosmological model $\Lambda$CDM, which takes into account the cosmic dynamics dominated by the cosmological constant ($\Lambda$) and cold dark matter (CDM).

Local estimates of $H_0$ are basically based on distance measurements with Cepheids and Type Ia Supernovae (SNe Ia). A recent estimate of $H_0$ based on this method is presented by A. G. Riess \textit{et al.} (2016)\cite{Riess2016}, where they obtained $H_0 = 73.24 \pm 1.74$ km s$^{-1}$ Mpc$^{-1}$ in $1 \sigma$ c.l., corresponding to a relative uncertainty of $2.4\%$ (including statistical and systematic uncertainties). The most important aspect of this method is the fact that all observables are obtained in low redshifts ($z \ll 1$) and, therefore, the values measured are almost completely independent of cosmological model.

The most current restrictive estimates of $H_0$ are obtained by global measurements which are based on Cosmic Microwave Background (CMB) observations, that is, they are measured in high redshifts ($z \sim 1000$). These measures are related to $H_0$ by assuming or inferring a cosmological model. Considering the standard flat $\Lambda$CDM model, Planck Collaboration (2016) \cite{PlanckCollaboration2016} obtained $H_0 = 66.93 \pm 0.62$ km s$^{-1}$ Mpc$^{-1}$ in $1 \sigma$ c.l., corresponding to a relative uncertainty of only $0.93\%$.

As the uncertainties associated with these two estimates are relatively small, the tension between them is approximately $3.4 \sigma$\cite{Feeney2017},
\begin{equation}
\label{3sidma}
\frac{\left | \widehat{H}_{0}^{Riess}-\widehat{H}_{0}^{Planck} \right |}{\left ( \sigma^{2}_{Riess}+\sigma^{2}_{Planck} \right )^{\frac{1}{2}}}\simeq 3.4.
\end{equation}
As well as previous tensions, for example, that of approximately $2.4 \sigma$ between the estimates of A. G. Riess \textit{et al.} (2011)\cite{Riess2011} and Planck Collaboration (2014)\cite{PlanckCollaboration2014}, whose obtained respectively $H_0=73.8 \pm 2.4$ e $67.3 \pm 1.2$ km s$^{-1}$ Mpc$^{-1}$ in $1 \sigma$ c.l., this latest of $3.4 \sigma$ has instigated several studies in order to define evidence of new physics or systematic errors hidden in the measurements of $H_{0}$. For having discarded the small possibility of coincidence, these measurements are the most frequent explanations for these tensions.

In this context, J. A. S. Lima and J. V. Cunha (2014)\cite{Lima2014} (from now on LC) proposed a new method capable to measure $H_0$ in intermediate redshifts ($z \approx 1$) in order to clarify a little bit more the problem mentioned before. The advantage of cosmological tests in intermediate redshifts is its independence from local calibrators \cite{Rephaeli1997}, as well as to be free from local and global effects, due to the distinct zone from the one where the anisotropies of the CMB are analyzed and the defined methods of the cosmic distance scale (Cepheids, SNE Ia, etc).

In LC determination, the authors used four different observables, they are i) angular diameter distance (ADD) to galaxy cluster based on the combination of Sunyaev-Zel'dovich effect and X-ray surface brightness of each galaxy cluster (SZE/X-ray technique), ii) ages of old high-redshift galaxies (OHRG), iii) observational measurements of the Hubble parameter ($H(z)$) and iv) baryon acoustic oscillations (BAO) peak. According to the authors, the cooperative interaction between these observables signicantly reduces the errors on $H_0 = 74.1 \pm 2.2$ km s$^{-1}$ Mpc$^{-1}$ in $1 \sigma$ c.l., that is, a relative uncertainty of approximately $3\%$. As the authors mentioned, this determination clearly favors local methods.

In their analysis, the authors (LC) used the ADD sample compiled by M. Bonamente \textit{et al.} (2006)\cite{Bonamente2006}, where the plasma distribution and dark matter were analyzed assuming the non-isothermal spherical double-$\beta$ model. However, the standard spherical geometry has been questioned due to the incompatibility with the so-called cosmic distance duality relation (CDDR) \cite{Holanda2010,Holanda2011,Meng2012,Holanda2012} and the observations from Chandra and XMM-Newton telescopes have shown that galaxy clusters exhibit a preferably elliptical X-ray surface brightness \cite{Fox2002,Jing2002,Limousin2013}.

However, the robustness of the results obtained by LC has already been questioned in the literature by R. F. L. Holanda \textit{et al.} (2014)\cite{Holanda2014}. Thus, these authors searched for possible hidden systematic errors by testing different hypotheses via the incubation time $t_{inc}$ (used in OHRG analysis), different cosmological models\footnote{The flat and non-flat LCDM model and the flat XCDM model.} and, besides the one used by LC, two other ADD samples compiled by E. D. Filippis \textit{et al.} (2005)\cite{De_Filippis2005}, the isothermal $\beta$-elliptical and isothermal $\beta$-spherical models, which describe the same galaxy cluster with different assumptions. They concluded that the estimated value for $H_0$ is weakly dependent on the cosmological models analyzed and the different hypotheses for the $t_{inc}$. However, even taking into account statistical and systematic errors, the authors found that the estimated value of $H_0$ varies considerably when the ADD sample is changed, which use different hypotheses about the properties of the galaxy clusters. For example, under the flat $\Lambda$CDM assumption for the model (the same cosmological model adopted by LC), R. F. L. Holanda \textit{et al.} (2014)\cite{Holanda2014} obtained $H_{0}=70.0_{-2.8}^{+3.0}$ and $65.0 \pm 3.0$ km s$^{-1}$ Mpc$^{-1}$ in $1\sigma$ c.l. for tests with the ADD sample whose cluster morphology was described by an isothermal $\beta$-elliptical model and an isothermal $\beta$-spherical model, respectively. In Appendix~\ref{Apendice1}, we present more information about these threes ADD sample and a brief discuss how the assumption of different gas distributions of the clusters can affect the measurement of ADD and, therefore, the estimates of $H_0$.

Therefore, a better understanding of the clusters morphology is necessary to transform the proposed determination by LC into a powerful tool to check $H_0$. In the absence of a better comprehension or consensus on the morphology of these clusters, it is natural to think about their exclusion from the estimation $H_0$. In this paper, we remove the ADD sample from the determination initially proposed by LC. Thus, the two main goals of this paper are: (i) to know if it has a relevant role under the results obtained by LC and (ii) to obtain a more accurate estimate for $H_0$ free from unknown systematic uncertainties. In addition, in order to obtain a more competitive estimate, we added new measurements of $H(z)$ and BAO peak, where we obtain $H_{0}=68.00 \pm 2.20$ ($68.71^{+1.37}_{-1.45}$) km s$^{-1}$ Mpc$^{-1}$ in $1 \sigma$ confidence level (c.l.) including statistical and systematic errors (statistical errors only), that is, we obtained results more consistent with global measurements of $H_0$.

This paper is organized as follows. In section~\ref{EqAmosMet} we present the basic equations, the samples and the methodology used in our analyses. In section~\ref{Resuldisc} we show the results and some discussions about them. Last but not least, in section~\ref{Conclusao} we present the conclusions.

\section{Basic equations, samples and methodology} \label{EqAmosMet}
\subsection{Basic equations}

In the flat $\Lambda$CDM model scenario, the Hubble parameter is usually expressed as
\begin{equation}
H(z)=H_{0}\sqrt{\Omega_{M}(1+z)^{3}+(1-\Omega_{M})}.
\label{Hz}
\end{equation}
where $\Omega_M$ is the current dimensionless parameter of matter density and $H_0$ the Hubble constant, generally expressed in terms of the dimensionless parameter $h\equiv H_{0}/(100$ km s$^{-1}$ Mpc$^{-1})$ . Assuming this model, it is easy to show that the age of the universe in a given redshift can be expressed by \cite{Alcaniz1999}
\begin{eqnarray}
t(z)&=&\frac{1}{H_{0}}\int_{0}^{1/(1+z)}\frac{dx}{x\sqrt{ \Omega_{M}x^{-3}+(1-\Omega_{M})}}\nonumber\\
&=& \frac{2}{3}\frac{H_{0}^{-1}}{\sqrt{1-\Omega_M}}\textrm{ln}\left[ \sqrt{\frac{(1-\Omega_M)}{\Omega_M}\left( \frac{1}{1+z}\right)^{3}} \right. \nonumber \\ 
&+& \left. \sqrt{\frac{(1-\Omega_M)}{\Omega_M}\left( \frac{1}{1+z}\right)^{3}+1} \right].
\label{t(z)}
\end{eqnarray}
The cosmological perturbations excited sound waves in the relativistic plasma in the early universe, such waves left printed a scale of preferred length in the photons and baryons distribution \cite{Peebles1970,Eisenstein2005} which, when analyzed through the powers spectrum of the baryonic matter, manifests itself as a series of oscillations \cite{Blake2011} called the baryons acoustic oscillations (BAO). The residual BAO peak can be described by a dimensionless parameter $A(z)$, defined by D. J. Eisenstein \textit{et al.} (2005)\cite{Eisenstein2005}, and for a flat universe, we obtain
\begin{eqnarray}
A(z)  \equiv  \frac{\sqrt{\Omega_{M}}}{E(z)^{1/3}}\left(\frac{1}{z}\int_{0}^{z}\frac{dz'}{E(z')} \right)^{\frac{2}{3}},
\label{Ateo} 
\end{eqnarray}
where $E(z)\equiv H(z)/H_{0}$ is the dimensionless Hubble parameter.

\subsection{Samples} \label{Amostras}

\paragraph{Hubble Parameter:} This sample consists of 40 observational measurements of $H(z)$ in the range of redshift $0.070 < z < 2.3$. Thus, all 40 measurements of $H(z)$ are presented in Table~\ref{Table1} related to each respective redshifts $z$, uncertainties $\sigma_{H(z)}$, references and methods used to obtain them. Among the 40 measures, the 18 measurements used by LC in their tests are indicated by the symbols $\dagger$.

\begin{table*}[!htb] %ph
	\caption{Hubble parameter measurements ($H(z)$) with their respect uncertainties ($\sigma_{H(z)})$, redshifts ($z$), references (Ref.) and the methods (Met.) used to obtain the analyses, which they are: the differential age of galaxies (DA), measurements of acoustic oscillations baryons peaks (BAO) and correlation function of luminous red galaxies (CF). Data from Ref.s with dagger $\dagger$ were used by LC. The dimensions of $H(z)$ and $\sigma_{H(z)}$ are in km s$^{-1}$ Mpc$^{-1}$.}
	\label{Table1}
	\centering
	\begin{tabular}{crrlc|crrlc}
		\hline\noalign{\smallskip}
		$z$	&$H(z)$& $\sigma_{H(z)}$ & Ref. & Met.&$z$ &$H(z)$& $\sigma_{H(z)}$ & Ref.&Met. \\
		\noalign{\smallskip}\hline\noalign{\smallskip}
		0.070&	69.0&	19.6& \cite{Zhang2014} & DA 		&0.480&	97.0&	62.0&\cite{Stern2010} $\dagger$ & DA\\
		0.090&	69.0&	12.0& \cite{Simon2005} $\dagger$ &DA		&0.510&	90.4&	1.9&\cite{Alam2016} &	BAO \\
		0.120&	68.6&	26.2& \cite{Zhang2014} &	DA		&0.570&	92.9&	7.8& \cite{Anderson2014}& BAO\\
		0.170&	83.0&	8.0& \cite{Simon2005} $\dagger$ &DA		&0.593&	104.0&	13.0&\cite{Moresco2012} $\dagger$ &DA	 \\
		0.179&	75.0&	4.0& \cite{Moresco2012} $\dagger$ &DA	&0.600&	87.9&	6.1&\cite{Blake2012} &	BAO \\
		0.199&	75.0&	5.0& \cite{Moresco2012} $\dagger$  & DA	&0.610&	97.3&	2.1 &\cite{Alam2016} & BAO \\
		0.200&  72.9&	29.6& \cite{Zhang2014} &DA			&0.679&	92.0&	8.0&\cite{Moresco2012} $\dagger$ &	DA  \\
		0.270&	77.0&	14.0& \cite{Simon2005} $\dagger$ &DA		&0.730&	97.3&	7.0&\cite{Blake2012} &	BAO \\
		0.280&	88.8&	36.6&\cite{Zhang2014}  &DA			&0.781&	105.0&	12.0&\cite{Moresco2012} $\dagger$ &DA	 \\
		0.350&	82.7&	8.4& \cite{Chuang2013} &	CF		&0.875&	125.0&	17.0&\cite{Moresco2012} $\dagger$ & DA\\
		0.352&	83.0&	14.0& \cite{Moresco2012} $\dagger$ & DA	&0.880&	90.0&	40.0&\cite{Stern2010} $\dagger$ &DA\\
		0.380&	81.5&	1.9& \cite{Alam2016} & BAO						&0.900&	117.0&	23.0&\cite{Simon2005}  & DA\\
		0.3802&	83.0&	13.5& \cite{Moresco2016}&DA						&1.037&	154.0&	20.0&\cite{Moresco2012} $\dagger$ &DA	 \\
		0.400&	95.0&	17.0& \cite{Simon2005} $\dagger$ &DA		&1.300&	168.0&	17.0&\cite{Simon2005} $\dagger$ &DA \\
		0.4004&	77.0&	10.2& \cite{Moresco2016}&DA						&1.363&	160.0&	33.6&\cite{Moresco2015}&DA	 \\
		0.4247&	87.1&	11.2& \cite{Moresco2016}&DA						&1.430&	177.0&	18.0&\cite{Simon2005} $\dagger$ & DA\\
		0.4400&	82.6&	7.8& \cite{Blake2012} &BAO			&1.530&	140.0&	14.0&\cite{Simon2005} $\dagger$ & DA\\
		0.4497&	92.8&	12.9&\cite{Moresco2016} &DA 					&1.750&	202.0&	40.0&\cite{Simon2005} $\dagger$ & DA\\
		0.470&	89.0&	34.0& \cite{Ratsimbazafy2017}&BAO				&1.9650& 186.5& 50.4& \cite{Moresco2015}& DA\\
		0.4783&	80.9&	9.0& \cite{Moresco2016}&DA						&2.300&	224.0&	8.0& \cite{Busca2013} &BAO \\
		\noalign{\smallskip}\hline			
	\end{tabular}  
    \end{table*}

\paragraph{Baryon acoustic oscillation peak:} This sample is composed by 4 measures. The first one is the same measure used by LC ($z = 0.35$) obtained from the search realized by Sloan Digital Sky Survey (SDSS), where $A(0.35)=0.469 \pm 0.017$ $(3.6\%)$\cite{Eisenstein2005}. The other three measures are the final set of WiggleZ Dark Energy Survey data, they are: $\textbf{A}^{ob}=(A(0.44),A(0.60),A(0.73))=(0.474,0.442,0.424)$ \cite{Blake2011}. According to C. Blake \textit{et al.} (2011)\cite{Blake2011} these last three measures are the most appropriate to be used in cosmological parameter estimates because, for the SDSS data, the value of $A(z)$ is obtained from the use of fiducial cosmological parameters and the same fractional error (for more details, see section 4.5 of Ref.~\cite{Eisenstein2005}). However, the chi-square statistic for the WiggleZ data in any cosmological model is obtained by multiplying the matrices $ (\textbf{A}^{ob}-\textbf{A}^{th})^{T} \textbf{C}^{-1}_{WiggleZ} (\textbf{A}^{ob}-\textbf{A}^{th})$, where the inverse covariance matrix $\textbf{C}^{-1}_{WiggleZ}$ is given by \cite{Blake2011}
\begin{equation}
\textbf{C}^{-1}_{WiggleZ}=\left( 
\begin{array}{rrr}  1040.3  & -807.5    & 336.8 \\ 
                    -807.5  & 3720.3    & -1551.9 \\
                    336.8   & -1551.9   & 2914.9 \end{array} \right).
\end{equation}

\paragraph{Ages of old high-redshift galaxies:} This sample consists of 11 measures of OHRG age in the range of redshift $0.62<z<1.8$ from selected subsamples of I. Ferreras \textit{et al.} (2009)\cite{Ferreras2009} and M. Longhetti \textit{et al.} (2007)\cite{Longhetti2007} samples. As mentioned by LC, this dataset is the one that provides the most accurate and restrictive ages (see Figure 1 of Ref.~\cite{Lima2014} for more details). This particular type of galaxy is interesting because we can assume an average incubation time with reasonable uncertainty, $t^{inc}=0.8 \pm 0.4$ Gyr \cite{Lima2009,Lima2014,Fowler1993,Sandage1993}, and thus estimate the age of the universe in different redshifts through the relation $t^{obs}_{i}=t^{gal}_{i}+t^{inc}$.

In some tests developed in this work, in addition to the statistical errors, we have also added systematic errors in quadrature. The technique used to date the OHRG is the comparison of the galaxy spectrum with the theoretical models of stellar population, whose systematic uncertainties, according to R. Jimenez \textit{et al.} (2004)\cite{Jimenez2004}, are not greater than $10\%-15\%$, since S. M. Percival \textit{et al.} (2009)\cite{Percival2009} considers uncertainties around $20\%$, as well as LC we adopted $15\%$ for OHRG measurements. On the other hand, we can observe in Table~\ref{Table1} that most of the measurements of $H(z)$ are obtained by the difference of age of galaxies (DA),
\begin{eqnarray}
H(z)=-\frac{1}{(1+z)}\frac{dz}{dt} \approx -\frac{1}{(1+z)}\frac{\Delta z}{\Delta t}.
\end{eqnarray}
And according to D. Stern \textit{et al.} (2010)\cite{Stern2010}, this method presents from $2\%$ to $3\%$ of systematic uncertainty. However, as done as LC we will be conservative and use $8\%$ for the $H(z)$ measures.

\subsection{Methodology}

The statistical analysis is performed by the construction of the $\chi^2$ function,
\begin{equation}
    \chi^{2}(\{\alpha\})=\sum^{N}_{i=1}\left[\frac{ F^{th}_{i}(\{\alpha\}) -F^{ob}_{i} }{\sigma_{i}}\right]^{2},
\end{equation}
where $F^{ob}_{i}$ represents the observational value with $\sigma_i$ being its respective uncertainty, $F^{th}_{i}$ is the corresponding theoretical predictions, $N$ is the total number of observational measurements and ${\alpha}$ is the set of free model parameters. From the $\chi^2$ function we are able to construct the probability density function (PDF),
\begin{equation}
    P(\{\alpha\})=Ae^{-\frac{1}{2}\chi^{2}(\{\alpha\})},
\end{equation}
where $A$ is the normalization factor. In the flat $\Lambda$CDM scenario, we have only two free parameters, $H_0$ and $\Omega_M$. Thus, as the parameter $H_0$ is the most interest parameter here, we generally make use of marginalization in $\Omega_M$,
\begin{equation}
    P(H_{0})=\int P(H_{0},\Omega_{M})d\Omega_{M}.
\end{equation}

In our first analysis, we estimated $H_0$ by using the following $\chi^{2}$,
\begin{eqnarray}
\label{x2dataLC}
    \chi^{2}(H_{0},\Omega_{M})&=&\sum_{i=1}^{18} \frac{\left[H^{th}_{i}(H_{0},\Omega_{M})-H^{ob}_{i}\right]^2}{\sigma^{2}_{i,est}+\sigma^{2}_{sys}} \nonumber \\
    &+&   \sum_{i=1}^{11} \frac{\left[t^{th}_{i}(H_{0},\Omega_{M})-(t^{gal}_{i}+t^{inc})\right]^2}{\sigma^{2}_{i,est}+\sigma^{2}_{inc}+\sigma^{2}_{sys}}  \nonumber \\
    &+&  \left[\frac{A^{th}(\Omega_{M})-0.469}{0.017}\right]^2 .
\end{eqnarray}
This is identical to the one proposed by LC, but except for the ADD data of galaxy clusters. Thus, we search to know if that sample has a relevant influence or role under the results found by LC.

On the other hand, for the second analysis, we use the following $\chi^2$
\begin{eqnarray}
    \label{x2newdata}
    \chi^{2}(H_{0},\Omega_{M}) &=& \sum_{i=1}^{40} \frac{\left[H^{th}_{i}(H_{0},\Omega_{M})-H^{ob}_{i}\right]^2}{\sigma^{2}_{est}+\sigma^{2}_{sys}} \nonumber \\ 
    & + &  \sum_{i=1}^{11} \frac{\left[t^{th}_{i}(H_{0},\Omega_{M})-(t^{gal}_{i}+t^{inc})\right]^2}{\sigma^{2}_{i,est}+\sigma^{2}_{inc}+\sigma^{2}_{sys}}  \nonumber \\
    & + & (\textbf{A}^{th}-\textbf{A}^{ob})^{T} \textbf{C}^{-1}_{WiggleZ} (\textbf{A}^{th}-\textbf{A}^{ob}) .
\end{eqnarray}
Here we include new $H(z)$ and the BAO peak measurements, but just like the previous one, we do not use the samples of ADD from galaxy clusters due to the reasons presented before. Therefore, we seek from this analysis to obtain a more competitive and accurate estimate for $H_0$ in intermediate redshift.

\section{Results and discussions} \label{Resuldisc}

\begin{figure}[!htb]
	\includegraphics[width=\columnwidth]{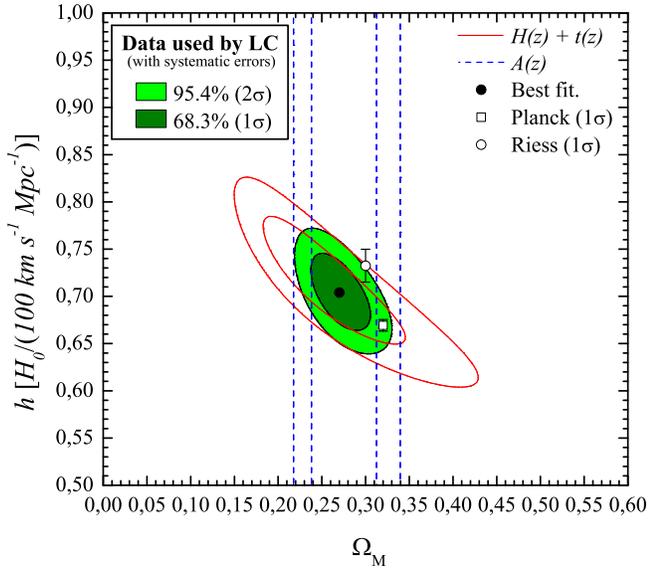}
	\caption{Contours throughout $1 \sigma$ and $2 \sigma$ c.l. of the parameters $h$ and $\Omega_M$ in the flat $\Lambda$CDM model obtained from the same samples (and with systematic errors) used by LC, but except for the samples of ADD from galaxy clusters. The contours filled in green colors refer to the joint analysis $H(z) + t(z) + A(z)$, where the best fit (black point) is $h = 0.7041$ and $\Omega_M = 0.270$. The white square and circle with error bars correspond to the estimates in $h$ in $1 \sigma$ c.l. obtained by Planck Collaboration (2016)\cite{PlanckCollaboration2016} and by A. G. Riess \textit{et al.} (2016)\cite{Riess2016}, respectively.}
	\label{Fig1}
\end{figure}

In Fig.~\ref{Fig1} and Fig.~\ref{Fig3}, it is shown the contours $1 \sigma$ and $2 \sigma$ c.l. of the $h$ and $\Omega_M$ parameters. Thus, the contours represented by dashed blue lines refer to the analysis using only $A(z)$, the ones represented by the solid red lines refer to the analysis using $H(z) + t(z)$ and, therefore, the contours filled in the green colors are referring to the joint analysis $H(z)+t(z)+A(z)$, where the best fit is represented by the black circle. In addition, the white square and circle with the error bars correspond to the estimates of $h$ in $1 \sigma$ c.l. obtained by Planck Collaboration (2016)\cite{PlanckCollaboration2016} and by A. G. Riess \textit{et al.} (2016)\cite{Riess2016}, respectively.

\begin{figure}[!htb]
	\includegraphics[width=\columnwidth]{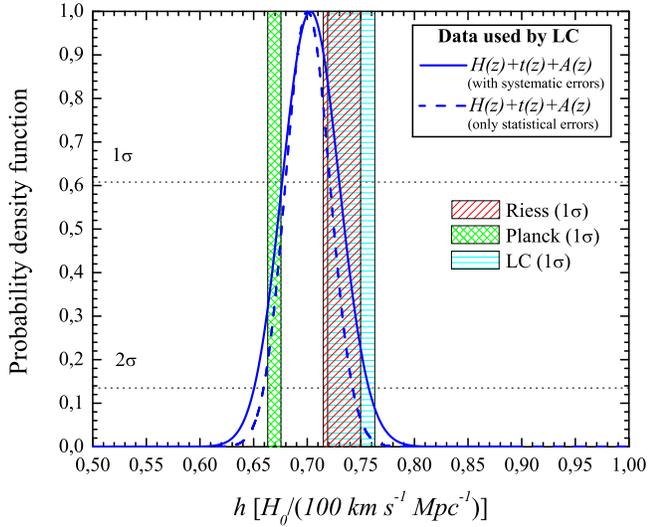}
	\caption{Marginal probability density function of the parameter $h$ only with statistical errors is presented by the dashed blue line and adding systematic erros, it is presented by the solid blue line obtained from the same samples used by LC, but except for the samples of ADD from galaxy clusters. The rectangles correspond to the values of $h$ in $1 \sigma$ c.l. obtained by LC, Planck Collaboration (2016)\cite{PlanckCollaboration2016} and by A. G. Riess \textit{et al.} (2016)\cite{Riess2016}. The two horizontal dotted lines delimit the regions of $1 \sigma$ and $2 \sigma$ confidence levels.}
	\label{Fig2}
\end{figure}

\begin{figure}[!htb]
	\includegraphics[width=\columnwidth]{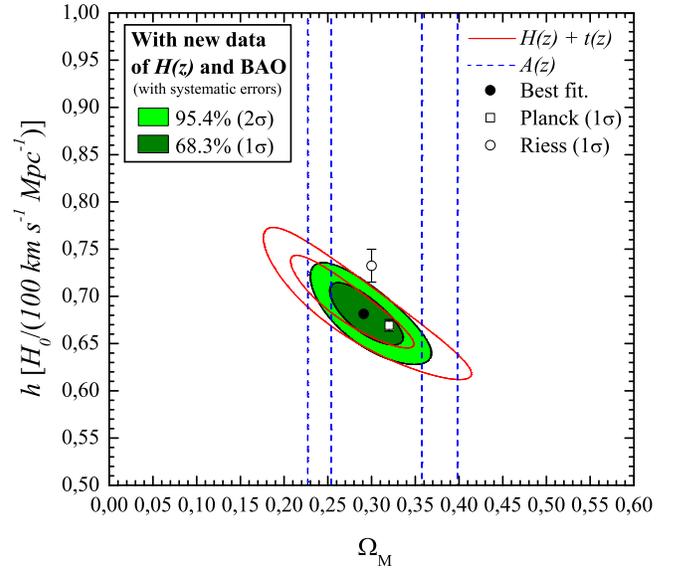}
	\caption{Contours throughout $1\sigma$ and $2\sigma$ c.l. of the parameters $h$ and $\Omega_M$ for the flat $\Lambda$CDM model obtained by adding the new $H(z)$ and BAO peak dataset. The contours filled in green colors refer to the joint analysis $H(z)+t(z)+A(z)$, where the best fit (black point) is $h = 0.6814$ and $\Omega_M = 0.291$. Thus, the white square and circle with the error bars correspond to the estimates of $h$ in $1 \sigma$ c.l. obtained by Planck Collaboration (2016)\cite{PlanckCollaboration2016} and by A. G. Riess \textit{et al.} (2016)\cite{Riess2016}, respectively. As shown in Fig.~\ref{Fig1}, the statistical and systematic errors have been also taken into account in the analyses.}
	\label{Fig3}
\end{figure}

The Fig.~\ref{Fig1} was produced by $\chi^2$ defined in Eq.~(\ref{x2dataLC}), where we used the same samples and systematic errors adopted by LC, but except for the ADD of galaxy clusters. For the joint analysis, we obtained $h = 0.7041^{+0.0410}_{-0.0410}$ $\left(0.7041^{+0.0685}_{-0.0660}\right)$ and $\Omega_{M}=0.270^{+0.036}_{-0.033}$ $\left(0.270^{+0.060}_{-0.052}\right)$ in $1\sigma$ ($2\sigma$) confidence levels. As shown in Fig.~\ref{Fig1}, this estimate no longer indicates any preference for global or local measurements in $H_0$, different from the result obtained by LC which clearly favors local measures. This statement is even more evident when we analyze the Fig.~\ref{Fig2}, where we exhibit our marginalized PDF results for the parameter $h$ with the values in $1\sigma$ c.l. obtained by LC, Planck Collaboration (2016)\cite{PlanckCollaboration2016} and by A. G. Riess \textit{et al.} (2016)\cite{Riess2016}. Including statistical and systematic errors (continuous blue line) in our analyses, we obtain $h=0.7030^{+0.0280}_{-0.0280}$ $\left(0.7030^{+0.0545}_{-0.0525}\right)$  with $\chi^2_{red} = 0.51$ and considering only with statistical errors (dashed blue line), we obtain $h=0.7003^{+0.0207}_{-0.0203}$ $\left(0.7003^{+0.0421}_{-0.0409}\right)$  with $\chi^{2}_{red} = 0.92$ in $1 \sigma$ ($2 \sigma$) confidence level.

\begin{figure}[!htb]
	\includegraphics[width=\columnwidth]{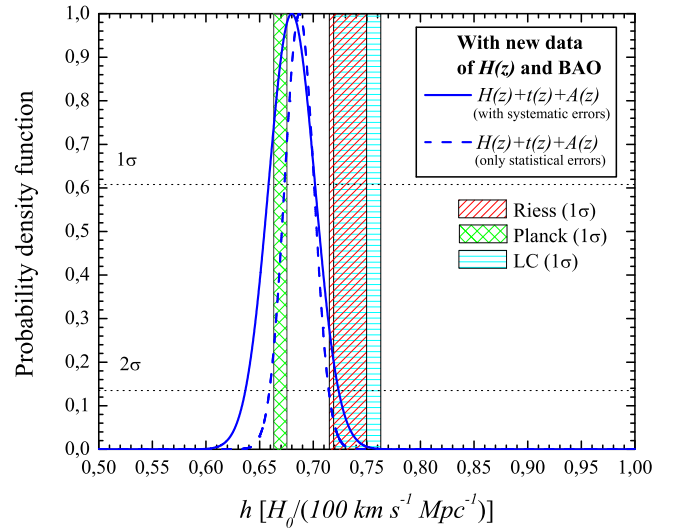}
	\caption{Marginal probability density function of the parameter $h$ only with statistical errors represented by dashed blue line. Including systematic errors is represented by solid blue line. Thus, these results were obtained by adding the new data of $H(z)$ and of the BAO peak. The dashed rectangles correspond to the values of $h$ in $1 \sigma$ c.l. obtained by LC, Planck Collaboration (2016)\cite{PlanckCollaboration2016} and by A. G. Riess \textit{et al.} (2016)\cite{Riess2016}. The two horizontal dotted lines delimit the regions of $1 \sigma$ and $2 \sigma$ confidence levels.}
	\label{Fig4}
\end{figure}

The Fig.~\ref{Fig3} was produced by using the $\chi^2$ from Eq.~(\ref{x2newdata}), where we added new measures of $H(z)$ and we also used three measurements of the parameter $A(z)$ from the WiggleZ final set data, in addition to the 11 OHRG data used by LC. For the joint analysis, Fig.~\ref{Fig3}, we obtain $h=0.6814^{+0.0334}_{-0.0332}$ $\left(0.6814^{+0.0541}_{-0.0538}\right)$ and $\Omega_{M}=0.291^{+0.044}_{-0.039}\left(0.291^{+0.076}_{-0.061}\right)$ in $1\sigma$ ($2\sigma$) confidence level. From these results, we might notice a small preference for the $H_0$ measurements obtained from global methods yet. Furthermore, the Fig.~\ref{Fig4} corroborates with this statement by showing the marginalized PDF of the parameter $h$ with the values $h$ in $1 \sigma$ c.l. obtained by LC, Planck Collaboration (2016)\cite{PlanckCollaboration2016} and by A. G. Riess \textit{et al.} (2016)\cite{Riess2016}. Here, we obtain the estimation with statistical and systematic errors (continuous blue line) $h=0.6800^{+0.0220}_{-0.0220}$ $\left(0.6800^{+0.0435}_{-0.0435}\right)$ with $\chi^{2}_{red.}=0.33$ and only with statistical errors (dashed blue line) we obatin $h=0.6871^{+0.0137}_{-0.0145}$ $\left(0.6871^{+0.0278}_{-0.0281}\right)$ with $\chi_{red.}^{2}=0.68$ in $1\sigma$ ($2\sigma$) confidence level. From Fig.~\ref{Fig4}, we might see that these estimates are in better agreement with the global measures of $H_0$ and with several other estimates of $H_0$ in intermediate redshifts, presented in Table~\ref{Table2}. Although, it does not agree with to the results obtained by LC.


\begin{figure}[!htb]
	\includegraphics[width=\columnwidth]{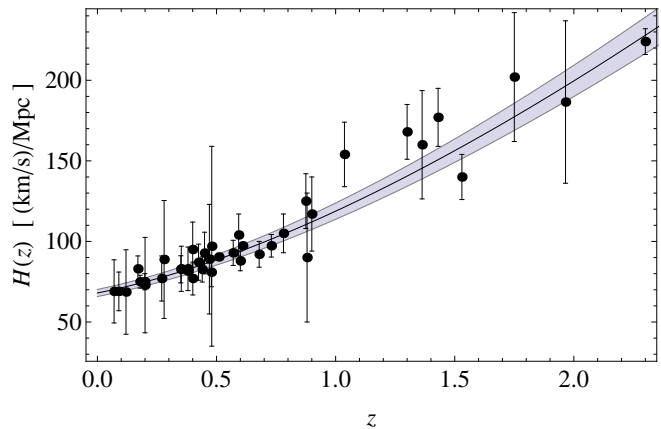}
	\caption{Hubble parameter as a function of $z$. The black solid line and the adjacent gray region represents the best fit and the region of $1 \sigma$ c.l. of $H(z)$ in the flat $\Lambda$CDM model (Eq.~(\ref{Hz})), assuming our estimate for $H_0$ ($68.00 \pm 2.20$) and $\Omega_{M}$ ($0.293 \pm 0.024$). The black dots with their respective error bars are the $H(z)$ measures shown in Table~\ref{Table1}. }
	\label{Fig5}
\end{figure}

\begin{figure}[!htb]
	\includegraphics[width=\columnwidth]{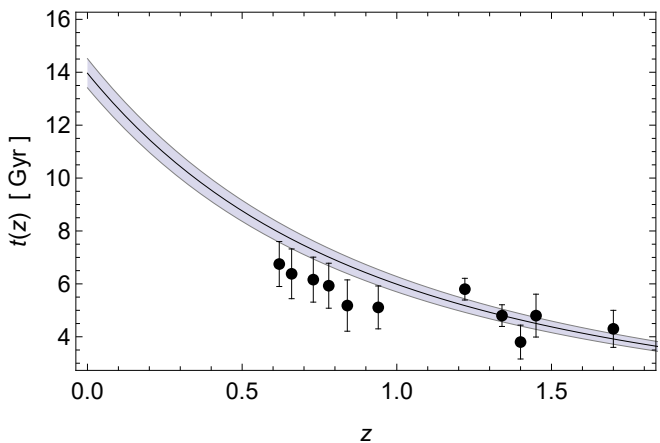}
	\caption{Age of the universe as a function of $z$. The black solid line and the adjacent gray region represents the best fit and the region of $1 \sigma$ c.l. of $t(z)$ in the flat $\Lambda$CDM model context (Eq.~(\ref{t(z)})) assuming our estimate for $H_0$ ($68.00 \pm 2.20$) and $\Omega_{M}$ ($0.293 \pm 0.024$). The black dots with their respective error bars are the measures of the age of the universe ($t^{obs}_{i}=t^{gal}_{i}+t^{inc}$) presented in Section~\ref{Amostras}.}
	\label{Fig6}
\end{figure}

From the $\chi^2$ function defined in Eq.~(\ref{x2newdata}), we might marginalize over $H_0$ and obtain the PDF value for $\Omega_M$, which results in $\Omega_{M}=0.293 \pm 0.024$ in $1\sigma$ c.l. with statistical and systematic errors. Therefore, using Eq.~(\ref{Hz}) and Eq.~(\ref{t(z)}) with the uncertainties propagation and the values here estimated for $H_0$ ($68.00 \pm 2.20$) and $\Omega_{M}$ ($0.293 \pm 0.024$), we obtain in $1\sigma$ c.l. the Hubble Parameter value and the age and universe in function of $z$, as well as shown in the Fig.~\ref{Fig5} and Fig.~\ref{Fig6}, respectively. Note that, the current age of the universe ($t_0$) is obtained by writing $z = 0$ in the Eq.~(\ref{t(z)}), so we might obtain $t_0 = 13.96 \pm 0.52$ Gyr in $1\sigma$ c.l. (shaded region of Fig.~\ref{Fig6} in $z = 0$). As well as expected, this value is very close to the one obtained by Planck Collaboration (2016)\cite{PlanckCollaboration2016}, which is $t_0=13.826 \pm 0.025$ Gyr in $1\sigma$ confidence level. Using the same process, we are able to estimate, among other parameters, the deceleration parameter ($q_0$), which is obtained by $q_0=3\Omega_M /2 -1 $ in the flat $\Lambda$CDM model \cite{Gival2018}, resulting $q_0 = -0.560 \pm 0.036$ $1\sigma$ c.l., which clearly indicates an accelerating expanding universe.

\section{Conclusion} \label{Conclusao}

In this work, we propose to investigate if the ADD clusters sample has a relevant influence on the results found by LC. In addition, we perform a competitive and more accurate estimate of $H_0$ in intermediate redshifts.

Excluding the samples of ADD from galaxy clusters from the determination proposed by LC, pointed in the literature as a source of systematic error, we obtain $H_0=70.30 \pm 2.80$ ($70.03^{+2.07}_{-2.03}$) km s$^{-1}$ Mpc$^{-1}$ in $1 \sigma$ c.l. including statistical and systematic errors (statistical errors only). Different from the results obtained by LC, this result does not indicate any preference for local or global measurements of $H_0$, since it is marginally compatible in the region of $1 \sigma$ c.l. with the one obtained by Planck Collaboration (2016)\cite{PlanckCollaboration2016} and by A. G. Riess \textit{et al.} (2016)\cite{Riess2016} (see Fig.~\ref{Fig2}). Thus, we might conclude that the ADD clusters sample used by LC has a significant influence on their results.

By aiming for a more accurate and competitive estimate, free of unknown systematic uncertainties, we improved the initial determination proposed by LC excluding the samples of ADD from galaxy clusters and adding new measurements of $H(z)$ and the BAO Peak. Thus, we obtained $H_0=68.00 \pm 2.20$ ($68.71^{+1.37}_{-1.45}$) km s$^{-1}$ Mpc$^{-1}$ in $1 \sigma$ c.l. including statistical and systematic errors (statistical errors only). These results are in full agreement with the results obtained by several other determinations $H_0$ in intermediate redshifts using different methods (see Table~\ref{Table2}). In addition, contrary to the results obtained by LC, ours clearly favor the global measurements of $H_0$ (see Fig.~\ref{Fig4} and Table~\ref{Table2}).

In addition, we marginalize $H_0$ to constrain $\Omega_M$, where obtained $\Omega_{M}=0.293 \pm 0.024$ in $1\sigma$ c.l. with statistical and systematic errors. Thus, using the flat $\Lambda$CDM model equations, uncertainties propagation and our estimates of $H_0$ and $\Omega_M$, we estimate the current age of the universe, $t_0 = 13.96 \pm 0.52$ Gyr, and the deceleration parameter, $q_0 = -0.560 \pm 0.036$, both in $1 \sigma$ confidence level. Finally, we have shown the good adjustment of the $H(z)$ and age of OHRG data with the Eq.~(\ref{Hz}) and Eq.~(\ref{t(z)}) assuming our estimate for $H_0$ and $\Omega_M$.

\begin{table*}[!hbt] %ph
\caption{Local, global and intermediate redshift determinations of $H_0$ in km s$^{-1}$ Mpc$^{-1}$.}
\label{Table2}
	\centering
\begin{tabular}{ccc}
\hline\noalign{\smallskip}
Determinations of $H_0$                                      & Local methods            & $H_0 \ (1\sigma)$   \\ 
\noalign{\smallskip}\hline\noalign{\smallskip}
Riess \textit{et al.} (2016)\cite{Riess2016}                &  SNe Ia/Cepheid           & $73.24^{+1.74}_{-1.74}$ \\
Freedman \textit{et al.} (2012)\cite{Freedman2012}          &SNe Ia/Cepheid             & $74.3^{+2.6}_{-2.6}$  \\
Riess \textit{et al.} (2011)\cite{Riess2011}                &  SNe Ia/Cepheid           & $73.8^{+2.4}_{-2.4}$ \\
LIGO Collaboration \textit{et al.} (2017)\cite{ligo2017gravitational}                        & GW170817                  &$70.0^{+12.0}_{-8.0}$ \\
\hline\noalign{\smallskip}
Determinations of $H_0$                                      & Global methods           & $H_0 \ (1\sigma)$   \\ 
\noalign{\smallskip}\hline\noalign{\smallskip}
Planck Collaboration (2016)\cite{PlanckCollaboration2016}   & CMB (Planck)              & $66.93^{+0.62}_{-0.62}$  \\
Planck Collaboration (2014)\cite{PlanckCollaboration2014}   & CMB (Planck)              & $67.3^{+1.2}_{-1.2}$  \\
Hinshaw \textit{et al.} (2013)\cite{WMAP2013}               & CMB (WMAP)                &  $70.0^{+2.2}_{-2.2}$  \\ 
\hline\noalign{\smallskip}
Determinations of $H_0$                                      & Intermediate methods    & $H_0 \ (1\sigma)$   \\ 
\noalign{\smallskip}\hline\noalign{\smallskip}
\textbf{LC} (2014)\cite{Lima2014}                        &$D_{A}+t(z)+H(z)^{\dagger}+A(0.35)$  & $74.1^{+2.2}_{-2.2}$ \\
\textbf{This work}: with new data                      & $t(z)+H(z)+\textbf{A}_{WiggleZ}$& $68.00^{+2.20}_{-2.20}$ \\
    T. M. C. Abbott \textit{et al.} (2017)\cite{Abbott2017} & DES + BAO + BBN  & $67.2^{+1.2}_{-1.0}$ \\
H. Yu \textit{et al.} (2017)\cite{Yu2017} &$H(z)$  & $67.0^{+4.0}_{-4.0}$ \\
Y. Chen \textit{et al.} (2017)\cite{Chen2017}                        &$H(z)$                         & $68.4^{+2.9}_{-2.6}$ \\
C. Cheng and Q. Huang (2015)\cite{Cheng2015} & BAO & $68.11^{+0.86}_{-0.86}$ \\
V. C. Busti \textit{et al.} (2014)\cite{Busti2014}                      & Param. reconst. of $H(z)$             & $64.9^{+4.2}_{-4.2}$ \\
\noalign{\smallskip}\hline
\end{tabular}
\end{table*}
%\footnotetext[2]{Combination Dark Energy Survey (DES) clustering and weak lensing data with Baryon Acoustic Oscillations (BAO) and Big Bang Nucleosynthesis (BBN).}

\appendix

\section{Samples of ADD from galaxy clusters} \label{Apendice1}

In this appendix, we shall present some important comments about the three samples of ADD from galaxy clusters obtained from the Sunyaev-Zel'dovich Effect and X-ray (the so-called ESZ / X-ray technique) observations and discuss how the assumption of different profiles of clusters can affect the measure of ADD and consequently the estimate of $H_0$.

\subsection{Isothermal spherical $\beta$-model and the isothermal elliptical $\beta$-model}

From a reanalysis of two samples, one with 7 \cite{Brian2001} and the other with 18 \cite{Erik2002} clusters which already had their measurements of ESZ and X-ray surface brightness collected, De Filippis \textit{et al.} (2005) \cite{De_Filippis2005} formed two 25 DDA samples in the range of redshift $0.023 < z < 0.784$, using two distinct models to describe the morphology of the same clusters, which are the isothermal elliptical $\beta$-model and the isothermal spherical $\beta$-model.

The isothermal spherical $\beta$-model is the simplest model existing in the literature, for this reason its main advantage of providing integrals with analytical solutions. For this model, the medium intraclusters (MIA) is considered isothermal and it is described by a spherical geometry with the following intracluster gas electronic density profile \cite{Cavaliere1978distribution,De_Filippis2005}:
\begin{eqnarray}
n_e(r)=n_{e0}\left( 1+\frac{r^2}{r_{c}^2}\right)^{-\frac{3}{2}\beta},
\label{ne(esf1)}
\end{eqnarray}
where $n_{e0}$ is the electronic density profile at the center of the MIA, $r$ is the radius from the center of the cluster, $r_c$ is the core radius of MIA and $\beta$ is the power law index.

In a reanalysis, De Filippis \textit{et al.} (2005)\cite{De_Filippis2005} made use of the isothermal elliptical $\beta$-model to describe the hot gas profile of MIA,
\begin{eqnarray}
n_e(r)=n_{e0}\left( 1+\frac{\theta_{1}^{2}+e^{2}_{proj}\theta_{2}^{2}}{\theta_{c,proj}^2}\right)^{-\frac{3}{2}\beta},
\label{ne(eli)}
\end{eqnarray}
where $e_{proj}$ is the axial ratio between the largest and the smallest axis projected in the plane of the sky, $\theta_{c,proj}$ is the angular radius of the core projected in the sky and $\theta_i$ is the angular coordinates that describe the projected positions.

\subsection{Non-isothermal spherical double-$\beta$ model}

Assuming the non-isothermal spherical double-$\beta$ model to describe the plasma distribution of the clusters, Bonamente \textit{et al.} (2006) \cite{Bonamente2006} compiled a sample composed by 38 galaxy clusters ADD in the range of redshift $0.14 < z < 0.89$. The function that describes the hot gas density in the MIA using this model is given by \cite{Bonamente2006}:
\begin{eqnarray}
n_e(r)=n_{e0} \left[ \frac{f}{\left( 1+\frac{r^2}{r_{c1}^2}\right)^{\frac{3}{2}\beta}} + \frac{(1-f)}{\left( 1+\frac{r^2}{r_{c2}^2}\right)^{\frac{3}{2}\beta}} \right],
\label{ne(duploesf)}
\end{eqnarray}
where $f$ is the fractional contribution of each portion ($0 \leq f \leq 1$), $r_{c1}$ and $r_{c2}$ are the two core radius which describe the shape of the inner and outer portions of the density distribution, respectively.

However, observations realized by Chandra and XMM-Newton Telescopes suggest that clusters do not have a spherically symmetric density profile, that is, they preferably exhibit elliptic surface brightness maps \cite{Fox2002,Sereno2006,De_Filippis2005,Limousin2013}. Another verification obtained from the XMM-Newton and Chandra telescopes is that the MIA is not isothermal \cite{Arnaud2009}. Thus, a more realistic model to describe the temperature profile of the MIA should be non-isothermal. In general, different gas profiles of the clusters do not affect the glow of the inferred central surface ($S_{X0}$) or the central Sunyaev-Zel'dovich decrement ($\Delta T_0$), but it gives different values for the core angular radius, $\theta_c$ (see Figure 1 of Ref.~\cite{De_Filippis2005}). A first order relationship between $\theta_{c,circ}$ and $\theta_{c,ell}$ was obtained by De Filippis \textit{et al.} (2005)\cite{De_Filippis2005}:
\begin{eqnarray}
\theta_{c,ell}=\frac{2e_{proj}}{1+e_{proj}}\theta_{c,circ},
\label{ne(elixesf)}
\end{eqnarray}
where $\theta_{c,ell}$ and $\theta_{c,circ}$ are the core angles obtained by the means of an isothermal elliptical $\beta$-model and an isothermal spherical isothermal $\beta$-model, respectively. This is an important detail because as the DDA is $\propto 1/H_0$ and $\propto 1/\theta_c$ (see Eq.(2.2) and Eq.(3.8) of the Ref.~\cite{Holanda2012H0}, respectively), different measurements of the core angular radius affect the ADD obtained by using the ESZ/X-ray technique and, consequently, the estimates of $H_0$. Therefore, the $H_0$ obtained by the spherical model is overestimated when compared to the one obtained by the elliptical model.

\acknowledgments

G. Pordeus da Silva thanks CNPq-Brazil and A. G. Cavalcanti thanks CAPES-Brazil for the financial support.

\medskip

\bibliography{mybibfile}

\end{document}